\documentclass[aps,pre,showpacs,twocolumn,superscriptaddress]{revtex4}
\usepackage{epsfig}
\usepackage{graphicx}
\usepackage{amsmath}
\usepackage{amssymb}
\usepackage{verbatim}   
\usepackage{color}      
\usepackage{subfigure}  

\newcommand{\beq}{\begin{equation}}
\newcommand{\eeq}{\end{equation}}

\begin{document}
\title{Reinforced walks in two and three dimensions}
\date{\today} 
\author{Jacob G. Foster} \affiliation{Complexity Science Group, Department of Physics and
  Astronomy, University of Calgary, Calgary, Canada} 
\author{Peter Grassberger} \affiliation{Complexity Science Group, Department of Physics and
  Astronomy, University of Calgary, Calgary, Canada}
  \affiliation{Institute for Biocomplexity and Bioinformatics,
  University of Calgary, Calgary, Canada } 
\author{Maya Paczuski} \affiliation{Complexity Science Group, Department of Physics and
  Astronomy, University of Calgary, Calgary, Canada}

\begin{abstract}
In probability theory, reinforced walks are random walks on a lattice (or more generally a graph) that preferentially revisit neighboring `locations' (sites or bonds) that have been visited before.  In this paper, we consider walks with one-step reinforcement, where one preferentially \emph{revisits} locations irrespective of the number of visits.  Previous numerical simulations [A. Ordemann {\it et al.}, Phys. Rev. E {\bf 64}, 046117 (2001)] suggested that the site model on the lattice shows a phase transition at finite reinforcement between a random-walk like and a collapsed phase, in both 2 and 3 dimensions. The very different mathematical structure of bond and site models might also suggest different phenomenology (critical properties, etc.).  We use high statistics simulations and heuristic arguments to suggest that site and bond reinforcement are in the same universality class, and that the purported phase transition in 2 dimensions actually occurs at zero coupling constant. We also show that a quasi-static approximation predicts the large time scaling of the end-to-end distance in the collapsed phase of both site and bond reinforcement models, in excellent agreement with simulation results.\end{abstract}

\maketitle

\section{Introduction}

\subsection{Background: Physics}
Random walks with memory have a large number of applications in physics and other sciences.
Many variants have thus been studied in different contexts. The best known example is 
presumably the self-avoiding walk \cite{deGennes}, which models the large scale behavior of flexible 
chain polymers in good solvents. As pointed out by Amit {\it et al.} \cite{Amit} the name 
`self-avoiding walk' is something of a misnomer, since this model describes either static self-avoiding 
{\it chains} or self-{\it killing} walks.  When the self-avoiding walker tries to revisit
a site it has visited before, it is not gently turned towards another neighboring site; it is killed.  In a more general version of this model, walkers carry an initial weight of unity, which decreases by a fixed factor whenever a site is revisited (the Domb-Joyce model \cite{domb}). If a site $i$ had been visited $n$ times before, the weight is diminished at the $n+1$ visit by $e^{nu}$ with $u<0$. 

When the sign of the interaction is changed to $u>0$, so that the weight is multiplied 
by a factor $e^{nu}>1$ at each revisit, the resulting self-attracting walk degenerates in any 
finite dimension; for large times, the walker just oscillates between two sites. This extreme
behavior is avoided if the weight change is independent of $n$ and one distinguishes only 
between sites which have and have not been visited before. This model is related to the 
Donsker-Varadhan \cite{donsker} ``Wiener sausage" problem \cite{mehra} and leads to 
a power law scaling $R_t \sim t^{1/(d+2)}$ for the end-to-end distance after 
$t$ time steps in $d$ dimensions of space.

In contrast to these ``static" models, where instances are weighted and the weights are modified by 
interactions, one can define ``dynamic" models where the walks are biased by the interaction. The oldest such model is the {\it true self-avoiding walk} (TSAW) of Amit {\it et al.} \cite{Amit}. Assume 
that at time $t$ the walker is at site $i$, and that the number of previous visits to any of the 
${\cal N}$ neighbors is $n_j,\; j=1,...,{\cal N}$. Then the probability to step to neighbor $j$ at 
the next time is 
\beq
    p_j = \frac{e^{n_ju}}{\sum_{j'=1}^{\cal N} e^{n_{j'}u}}\quad u < 0\;.     \label{TSAW}
\eeq
This is a much milder modification than the original self-avoiding walk. Accordingly, the 
r.m.s. end-to-end distance scales as $R^2_t \sim t$ for $d>2$, while there are logarithmic 
corrections at the `upper critical dimension' $d=d_c=2$. In contrast, the upper critical dimension
for the self-avoiding walk is $d_c=4$, and $R^2_t \sim t^{2\nu_d}$ for $d<d_c$ with $\nu_d<1/2$
\cite{deGennes}.

\subsection{True self-attracting walks}

When the sign of $u$ is switched to positive, the resulting ``true self-attracting walks"
(TSATWs) are also closer to random walks than the ordinary self-attracting walks. It seems that the 
behavior of the TSATW with $p_j$ given by Eq.(\ref{TSAW}) but with $u>0$ is unknown. On the other 
hand, there are several numerical studies of the TSATW with one-step reinforcement 
\cite{sapo,reis,prasad,lee,ordemann1,ordemann2}
\beq
    p_j = \frac{e^{\kappa_ju}}{\sum_{j'=1}^{\cal N} e^{\kappa_{j'}u}}\quad u > 0\;.     \label{TSAW1}
\eeq
where $\kappa_j=0$ if the site $j$ has never been visited before, and $\kappa_j=1$ otherwise. By far 
the most extensive studies were those of \cite{ordemann1,ordemann2}, which claimed that one-step reinforcement TSATWs on the lattice showed a non-trivial phase transition in both $d=2$ and $d=3$, with $u_c(d=2) = 0.88\pm 0.05$ and $u_c(d=3) = 1.92\pm 0.03$. In both cases, the behavior of $R^2_t$ is supposed to change at $u_c$ from $R^2_t\sim t$ at $u<u_c$ to 
\beq 
   R^2_t\sim t^{2/(d+1)}                 \label{R2}
\eeq
 at $u>u_c$.  Hence the phase transition is between a random-walk-like and a collapsed phase.  At the critical point, $R_t$ scales with a new exponent $\nu_c$ which is $0.40\pm 0.01$ in $d = 2$ and $0.303\pm 0.005$ in $d = 3$ \cite{ordemann2}. These phase transitions are also seen in the average number $\langle S_t \rangle$ of sites visited up to time $t$. This scales as $\langle S_t\rangle \sim t$ for $u<u_c$ (with a logarithmic correction for $d = 2$), but as $t^k$ with 
\beq
   k=d/(d+1)
\eeq
for $u>u_c$. The latter was derived from a quasi-static approximation \cite{dalmaroni} in 
Ref.~\cite{sapo}.  The quasi-static approximation seems to be satisfied to high precision (see below). At $u=u_c$, Ordemann {\it et al.} found $k_c = 0.80\pm 0.01$ ($d = 2$) resp. $0.91\pm 0.01$ ($d = 3$) \cite{ordemann2}.

\subsection{Background: Mathematics}

In a parallel and largely independent development, these and similar random walks with 
memory have been extensively studied in the probability theory literature.  For a recent survey, see 
\cite{pemantle}.  The rigorous mathematical study of reinforced walks displays much more breadth than the rather limited study of one-step site reinforcement in the statistical physics literature.  In contrast to the physics literature, which focuses on the site model, {\it bond} or ``edge" reinforced random walks (ERRW) have been studied in great detail and with multiple reinforcement as in Eq.(\ref{TSAW}) with positive $u$.  Such walks (most clearly on trees) are closely related to P\'{o}lya urn processes and similar problems with reinforcement that can be solved exactly (note that walks with bond reinforcement are often called `trails' in the physics literature \cite{trails}).  For the models with multiple \emph{site} reinforcement discussed above (called vertex-reinforced random walks or VRRW), the related urn process is Friedman-like and less tractable \cite{pemantle, pemantle1}, see endnote \footnote{We summarize the difference between these two urn processes following \cite{pemantle1}.  In the simplest P\'{o}lya urn, one begins with a single red and black ball.  In a series of draws, one removes a ball at random and returns it to the urn with another of the same type.  The fraction of red balls, surprisingly, converges to a random limit uniformly distributed on the closed interval $[0,1]$.  In the Friedman urn, one reinforces both the drawn color and the other color.  More surprisingly, the limit here converges to {1/2} even if the reinforcements are extremely biased towards the drawn color -- as in \cite{pemantle1} $1000$ to $1$ -- although the convergence is extraordinarily slow.}.

The mathematicians have discovered profound differences between these two models.  For example, ERRW is recurrent on finite graphs \cite{dia88, kr99}, meaning that every edge is traversed infinitely often, while VRRW is not, becoming trapped e.g. on a line of five vertices \cite{pemantle, pemantle1, tar04} or more generally on ``trapping subgraphs" \cite{pemantle, pemantle1, pv99, vol1}.  Many properties of the ERRW remain unknown; for example, the recurrence of ERRW on the infinite 2-d lattice is an open question.  Even the one-step ERRW model (called once-reinforced in the mathematics literature) has only been successfully studied on a few special graphs, e.g. the infinite regular or Galton-Watson tree (where it is transient \cite{pemantle, dkl02, die05}) or the infinite ladder (where it is recurrent \cite{pemantle, sel06}).  The recurrence of one-step ERRW on the infinite 2-d lattice remains essentially open, although Sellke showed the separate recurrence of each coordinate \cite{pemantle, sellke}.  Pemantle \cite{pemantle, pemantle1} provides a more complete description of these and other results. 

The difference in mathematical tractability and underlying structure might suggest that models with bond and site reinforcement show different phenomenology. But this result would be unexpected from considerations of universality in statistical physics.

\subsection{Overview of Results}

In the present paper, we clarify some of these issues by means of high precision 
simulations. Our main results are:
\begin{itemize} 
\item Bond and site reinforced TSATWs with one-step reinforcement show the same critical behavior and are likely in the same universality class;
\item There is no finite $u$ phase transition in the 2-dimensional TSATW model with one-step
reinforcement.  Walks are in the collapsed phase for all $u>0$ and the phase 
transition happens at $u_c=0$;
\item The critical point and the critical exponents for TSATWs with one-step reinforcement in
$d = 3$ are markedly different from the values obtained in \cite{ordemann1,ordemann2}; and
\item The quasistatic approximation for the end-to-end distance seems to become exact as $t\to\infty$ for the collapsed
phase.
\end{itemize}

\section{Numerical Methods and Results}

\subsection{Methods}
\label{methods}

Simulations of TSATWs with one-step reinforcement are straightforward. To keep track of 
previous visits, one has to store a one-bit ``spin" variable $s_i$ for each site (bond) $i$, and 
clear all spins after each walk. For convenience we sometimes used one byte per spin, which has the 
added advantage that clearing is needed only after every 255th walk. This requires 
$L^d/8$ resp. $L^d$ bytes of memory for site TSATWs and $dL^d/8$ resp $dL^d$ bytes for bond 
TSATWs. Memory limitations were 
more severe than CPU time so in the following we show more detailed results for site TSATWs than for bond TSATWs.

The most serious potential source of systematic errors arises from lattices that are too small. If 
open boundary conditions (b.c.) are used, the walk cannot go beyond the boundary, and both $R_t$ and $\langle S_t\rangle$ are underestimated. If periodic b.c. are used and the walk wraps around the lattice, it finds visited terrain in front of it and $R_t$ is overestimated, while $\langle S_t\rangle$ is still
underestimated.
We used lattices with helical boundary conditions and with up to $N = 2^{32}$ sites ($d = 2$)
resp. $2^{34}$ sites ($d = 3$),
see endnote \footnote{We always used lattice sizes such that the total number $N$ of sites was a power of 
two, in order to implement the boundary condition $i\equiv i-N\;\mod N$ by bit masking. The 
requirement that $N=L^d$ was not strictly enforced (unless $N$ was such that this would 
give an integer $L$), so the lattices were only approximately of cubic shape. This is of 
no consequence in the following.}. For each walk, the spans $x_{\rm max}-x_{\rm min}$ in all 
$d$ directions were measured, and it was checked that the fraction of walks where any span was 
$\ge L$ did not exceed $10^{-4}$. This restricted the number of steps per 
walk to $t_{\rm max} \le 10^8$ for $d = 2$, and to $t_{\rm max} \approx 4\times 10^7$ for $d = 3$. The total number 
of walks for each parameter setting was typically $\approx 2\times 10^4$ to $\approx 2\times 10^5$.  

\begin{figure}
  \begin{center}
  \epsfig{file=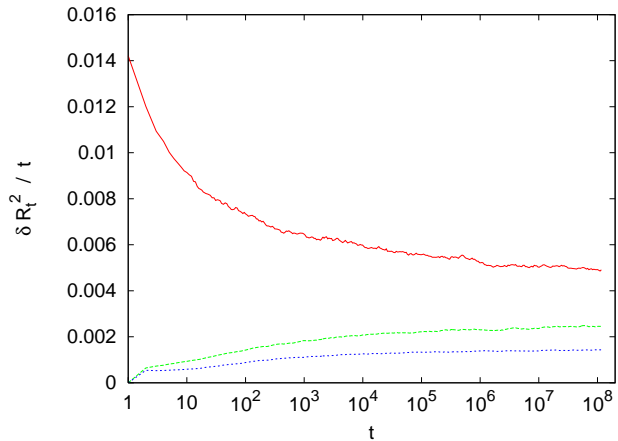, width=6.cm, angle=270}
  \caption{(color online) Statistical errors (1 st.d.) of the three estimators for the average 
      squared end-to-end distance divided by $t$ (direct, $\widehat{R_t^2}/t$, and 
      $[R_t^2]_{\rm opt}/t$ from top to bottom) for 2-d site TSATWs with $u=0.34$. The number of 
      walks in this sample was $10^4$, which is only a small fraction of our total sample.}
  \label{fig2var}
  \end{center}
\end{figure}

\subsection{Variance reduction}
\label{variance}
For small $u$, where walk-to-walk variation is significant, substantially increased accuracy is obtained by the following variance reduction procedure. Assume that the walker has already made $t$ steps and is presently at a site with cartesian coordinates ${\bf x}_t$. Given ${\bf x}_t$ and the states of the neighboring sites (i.e., visited or unvisited), one can calculate the expected increment $\widehat{\Delta \bf x}_{t+1}$ for the next step, since one knows the probability for the walker to step in each direction. From this, one obtains an estimate for the increment of $R_{t+1}^2$ 
\beq
    \widehat{\Delta R_{t+1}^2} = \widehat{[{\bf x}_{t+1}^2 - {\bf x}_{t}^2]} = 2 {\bf x}_t \cdot \widehat{\Delta \bf x}_{t+1} +1\;.
\eeq
where we have used the fact that $\widehat{ \Delta \textbf{x}_{t+1} \cdot \Delta \textbf{x}_{t+1}}  = 1$.  The improved estimate is obtained by summing these increments,
\beq
    \widehat{R_{t}^2} = \sum_{t'=1}^t \widehat{\Delta R_{t'}^2}\;.
\eeq
Further improvement is obtained by taking the optimal linear combination of the direct sample 
average and this estimator,
\beq
    [R_t^2]_{\rm opt} = \alpha_t R_t^2 + (1-\alpha_t) \widehat{R_t^2} \;,
\eeq
with $\alpha_t$ fixed for each $t$ such that the variance of $[R_t^2]_{\rm opt}$ is minimal.  Differentiating  the variance of $[R_t^2]_{\rm opt}$ with respect to $\alpha_t$ and minimizing this variance requires the estimation of the variances of $R_t^2$ and $\widehat{R_t^2}$ as well as their covariance. In Fig.~\ref{fig2var} we show the errors (single standard deviations, divided 
by $t$) of the three estimators for 2-d site TSATWs with $u=0.34$.

\subsection{Site TSATWs in d = 3}
\label{site3}
\begin{figure}
  \begin{center}
  \epsfig{file=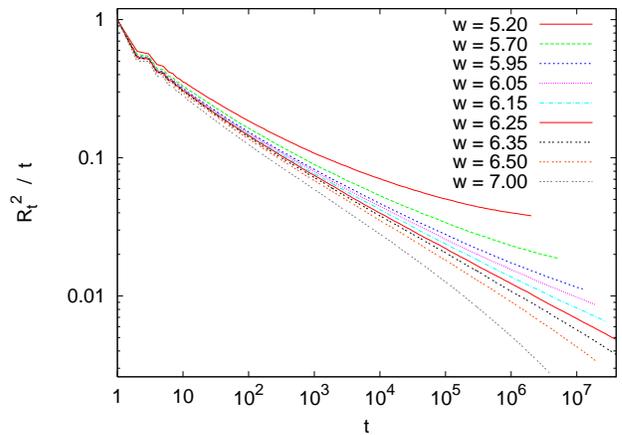, width=6.cm, angle=270}
  \caption{(color online) Average squared end-to-end distances for site TSATWs in 3 dimensions, divided 
      by the number $t$ of steps. Each curve corresponds to a fixed value of $u$, with $u$ increasing 
      from top to bottom. Here and in all subsequent figures, the curves are labelled by $w = \exp(u)$.
      The critical value $u=u_c$ corresponds to a straight curve in the limit 
      $t\to\infty$ whose slope is $2\nu_c - 1$. Statistical errors are comparable to the thickness
      of the curves.}
  \label{figR3}
  \end{center}
\end{figure}

The \emph{average} r.m.s end-to-end distance divided by the number of steps, $t^{-1}R_t^2$, is shown 
in Fig.~\ref{figR3} for site TSATWs in $d = 3$. In this and in all subsequent figures, curves are 
not labelled by $u$ but by $w = \exp(u)$. We see clearly that there are significant corrections
to scaling (all curves bend upward at small $t$), but they are no worse than in other 
nonequilibrium critical phenomena. A more careful analysis, taking these corrections into 
account, gives $u_c = 1.831\pm 0.002$ ($w_c = \exp(u_c) = 6.24\pm 0.01$) and $\nu_c = 0.378\pm 0.004$. In particular, we can rule 
out the possibility that $u_c > 1.85$ from the simple fact that all curves for $u > 1.85$ (i.e. for $w > 6.35)$ are clearly S-shaped and curve down at large $t$. These estimates are incompatible
with those of \cite{ordemann2}, $u_c = 1.92\pm 0.03$ and $\nu_c = 0.303\pm 0.005$. Possible
explanations for these earlier results are that corrections to scaling were neglected in \cite{ordemann2} or that the lattices used were too small.

\begin{figure}
  \begin{center}
  \epsfig{file=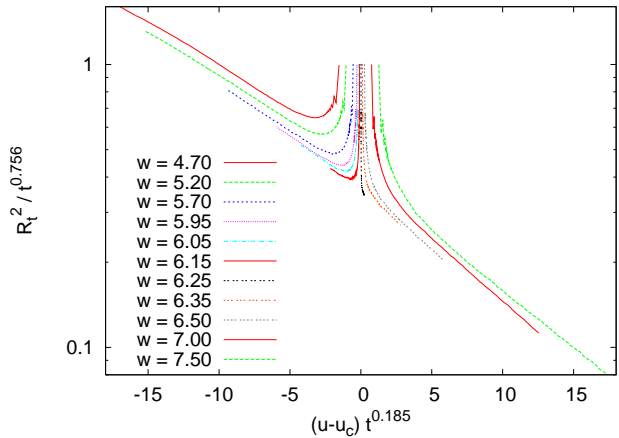, width=6.cm, angle=270}
  \caption{(color online) The same data as in Fig.~\ref{figR3}, plotted now as $R_t^2/t^{2\nu_c}$ 
      versus $(u-u_c)t^\phi$, with $\nu_c = 0.378, u_c = 1.831$ $(w_c = 6.24)$, and $\phi=0.185$. If there were 
      perfect scaling, all data would fall on a single curve.}
  \label{figcoll3}
  \end{center}
\end{figure}

The cross-over behavior near $u\approx u_c$ can be fitted to the usual ansatz
\beq
    R_t^2 = t^{2\nu_c} F[(u-u_c)t^\phi] \;\;+\;{\rm corrections}        \label{scale3}
\eeq
with $\phi=0.185\pm 0.020$, as seen from the data collapse shown in Fig.~\ref{figcoll3}.
The deviations from a perfect collapse seen in this figure are due to the corrections to scaling at small $t$ seen in Fig.~\ref{figR3}, which are not included in the scaling ansatz Eq.~(\ref{scale3}). 
The apparent collapse could have improved by the widespread practice of plotting the sub- and 
supercritical branches separately, without demanding that they join smoothly (the function $F(z)$
must be analytic at $z=0$). But the results obtained in this way would be spurious.

\begin{figure}
  \begin{center}
  \epsfig{file=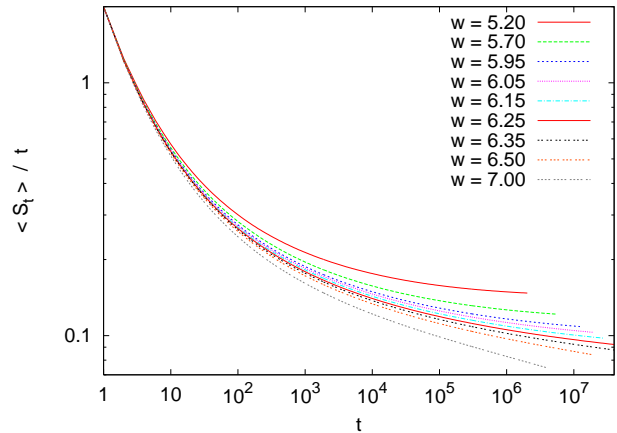, width=6.cm, angle=270}
  \caption{(color online) Average number of visited sites for site TSATWs in 3 dimensions, divided by
      the number $t$ of steps. Curves correspond to the same values of $u$ as in Fig.~\ref{figR3}.
      Statistical errors are much smaller than the thickness of the curves.}
  \label{figS3}
  \end{center}
\end{figure}

Results for the average number of visited sites, $\langle S_t\rangle$, again divided by $t$, are 
shown in Fig.~\ref{figS3}. This time the corrections to scaling are much bigger. This is not 
unexpected, since there are also large corrections to the asymptotic law $\langle S_t\rangle \sim t$
for ordinary 3-d random walks. These corrections make an independent estimate of $u_c$ impossible, 
whence we shall use the estimate obtained from $R_t$, i.e. $u_c = 1.831\pm 0.002$. The corrections to scaling also make the estimation of the exponent $k_c$ very 
uncertain, in spite of the extremely small statistical errors (much smaller than the thickness of the curves).  Our best estimate is $k_c = 0.977\pm 0.010$.  This is again incompatible with the estimate $0.91\pm 0.01$ of \cite{ordemann2}. The leading correction to scaling exponent, defined as $\langle S_t\rangle = t^{k_c} [a +b/t^\Delta + o(t^{-\Delta})]$, is found to be $\Delta =0.22\pm 0.03$. This is to be compared to $\Delta =1/2$ for ordinary 3-d walks \cite{torney}.
 
For the supercritical case, $u>u_c$, the following argument was given in \cite{sapo}: Let us assume that 
the visited sites form, for large $t$, a compact $d-$dimensional domain $V_t$ whose volume increases as 
$S_t \equiv |V_t| \propto R_t^d$ with $R_t \sim t^\nu$. Its surface is fuzzy but not fractal, i.e. it 
increases as $|\partial V_t| \propto R_t^{d-1}$.  If the walker is uniformly distributed inside $V_t$, 
then the chance for it to be at the boundary is $|\partial V_t| / S_t \propto 1/R_t$. This is then also 
proportional to the chance that the walker will make the next step outside $V_t$, i.e. $d\langle S\rangle
/dt \sim t^{-\nu}$. Integrating this gives $\nu = 1/(d+1)$ \cite{sapo}. The main assumption here 
is \emph{not} that $\partial V_t$ is non-fractal (as stated in \cite{sapo,ordemann2}), but that the 
walker is uniformly distributed inside $V_t$. This would be exact if the boundary would not grow at all
(i.e. in the limit $u\to\infty$), but for finite $u$ it corresponds to a {\it quasistatic 
approximation} in the sense of \cite{dalmaroni}. 

\begin{figure}
  \begin{center}
  \epsfig{file=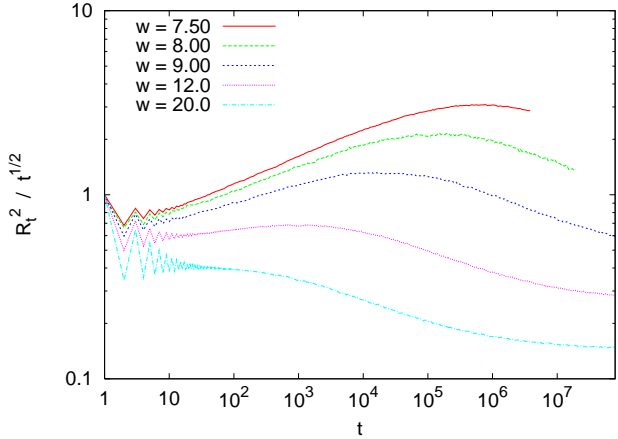, width=6.cm, angle=270}
  \caption{(color online) $R_t^2 / \sqrt{t}$ for site TSATWs in 3 dimensions, for five values of 
      $u$ which are all larger than $u_c$. The curves seem to become horizontal as $t\to\infty$, with the asymptotic behavior appearing for smaller values of $t$ as $u$ increases.}
  \label{figR3s}
  \end{center}
\end{figure}

In order to test this quasistatic approximation of the supercritical behavior, we plot in Fig.~\ref{figR3s} the ratio $R_t^2/\sqrt{t}$ for several values of $u>u_c$. We see very large corrections to scaling (the corrections to $\langle S_t\rangle$ would be even larger), but the curves do seem to become horizontal for $t\to\infty$. For $u\ge 2.5$, our best estimate is $R_t\sim t^\nu$ with $\nu =0.25\pm 0.01$, in perfect agreement with the prediction of the quasistatic approximation.

\subsection{Site TSATWs in $d = 2$}
\label{site2}
In two dimensions the situation seems at first glance similar, except for the fact that 
corrections to scaling are even larger. The latter is not surprising: random walks are recurrent 
in $d = 2$, while they are not in any $d > 2$. The number of visited sites increases not as $t$
in $d = 2$, but as $S_t = \pi t/\ln(8t)[1+O(1/\ln t)]$ \cite{torney}. Related to this is the fact that 
true self avoiding walks have upper critical dimension $d = 2$, leading to logarithmic corrections in 
most observables for $d = 2$. As a consequence, one should also expect logarithmic corrections for 
TSATWs.

\begin{figure}
  \begin{center}
  \epsfig{file=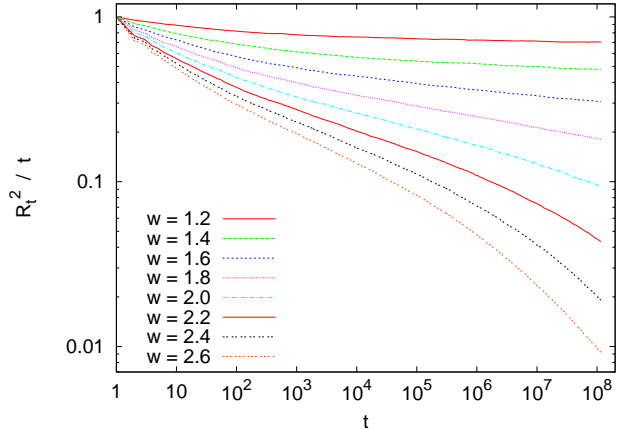, width=6.cm, angle=270}
  \caption{(color online) Same as Fig.~\ref{figR3}, but for $d = 2$.}
  \label{figR2}
  \end{center}
\end{figure}

Results for the end-to-end distance are shown in Fig.~\ref{figR2}. Again we show a log-log plot 
of $R_t^2/t$, for easy comparison with Fig.~\ref{figR3}. The main difference between these two 
plots is that the curves fan out in Fig.~\ref{figR2} already for very small $t$, while they fan 
out only at much later times in Fig.~\ref{figR3}. While the curves for $u<u_c$ in Fig.~\ref{figR3}
first seem to follow the scaling $R_t \sim t^{\nu_c}$ and cross over to $R_t \sim t$ only at 
large $t$, no such cross-over is seen in Fig.~\ref{figR2}. Careful inspection shows that all 
curves for $u>0.58$ (i.e. $e^u > 1.79$) bend down at large $t$, indicating that $u_c \le 0.58$ and that the estimate
$u_c=0.88\pm 0.05$ of \cite{ordemann2} is untenable. If we want to see a critical point with an
associated non-trivial power law in these data, then a possible candidate is $u_c\approx 0.54$ and 
$\nu_c \approx 0.47$.

\begin{figure}
  \begin{center}
  \epsfig{file=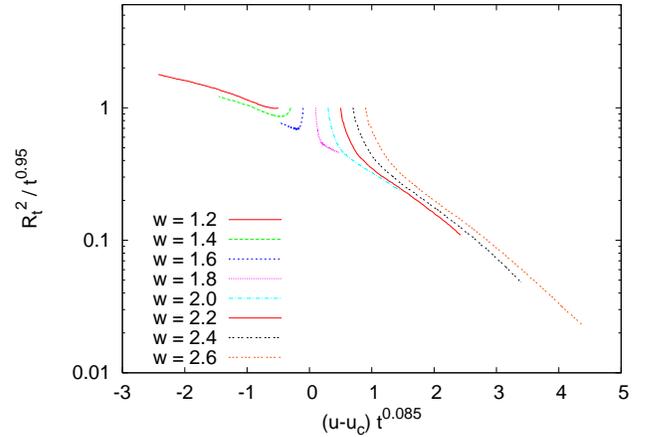, width=6.cm, angle=270}
  \epsfig{file=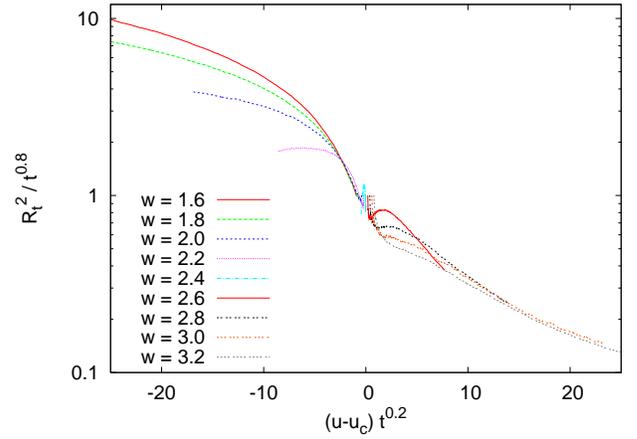, width=6.cm, angle=270}
  \caption{(color online) Attempted data collapse analogous to Fig.~\ref{figcoll3}, but for the 
      data of Fig.~\ref{figR2}. Parameters in panel (a) are $\nu_c = 0.475, u_c = 0.548$ $(w_c = 1.730)$, and 
      $\phi=0.085$, while panel (b) uses the values $\nu_c = 0.40, u_c = 0.88$ $(w_c = 2.411), \phi=0.2$ 
      proposed in \cite{ordemann2}.} 
  \label{figcoll2}
  \end{center}
\end{figure}

An attempted data collapse for the data of Fig.~\ref{figR2}, again using Eq.(\ref{scale3}) and 
optimized values $u_c = 0.548, \nu_c = 0.475$, and $\phi = 0.085$, is shown in Fig.~\ref{figcoll2}.
We might mention that the exponents proposed in \cite{ordemann2}, $\nu_c = 0.40\pm 0.01$ and 
$\phi \approx 0.2$, seem to be ruled out. A data collapse using these exponents is shown in
panel (b) of Fig.~\ref{figcoll2}. Although it has an acceptable overall dispersion, this is 
achieved mainly by fitting well the small-$t$ data, and grossly misrepresenting data for 
large $t$.

\begin{figure}
  \begin{center}
  \epsfig{file=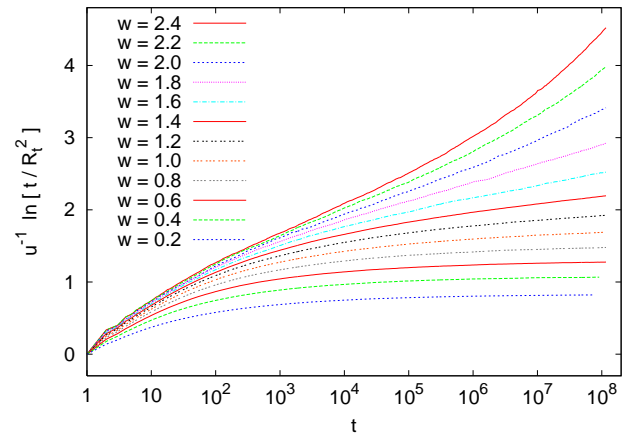, width=6.cm, angle=270}
  \caption{(color online) The function $\Psi_t(u)$ [see Eq.~(\ref{psi})] plotted against $t$ for
      both positive and negative values of $u$, including $u=0$ $(w = 1)$. The fact that this figure resembles 
      a typical cross-over plot as in Fig.~\ref{figR3} suggests that $u=0$ is a critical point.}
  \label{figPsi}
  \end{center}
\end{figure}

\begin{figure}
  \begin{center}
  \epsfig{file=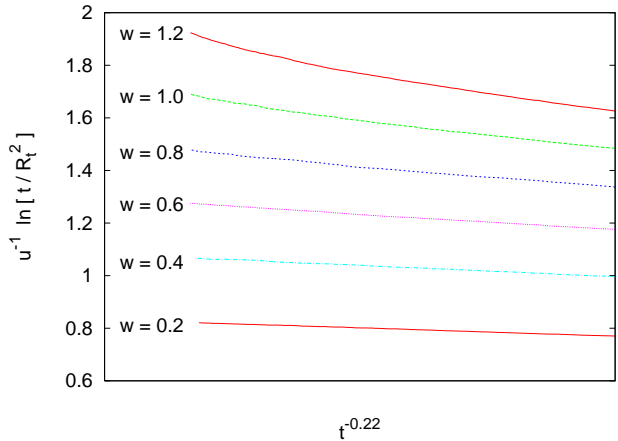, width=6.cm, angle=270}
  \caption{(color online) Part of the data shown in Fig.~\ref{figPsi}, but plotted against $1/t^{0.22}$. 
      This exponent gave the straightest curves for $w= \exp(u) <0.5$. No exponent would give straight 
      curves for $w\geq 0.8$.}
  \label{figPsi2}
  \end{center}
\end{figure}

Although the collapse seen in Fig.~\ref{figcoll2}(a) is satisfactory, the smallness 
of $\phi$ and the closeness of $\nu_c$ to the random walk value $\nu=1/2$ suggest a very different 
interpretation. We propose that there is in fact \emph{no} phase transition at any $u_c>0$. Instead, 
the TSATW is collapsed for any $u>0$, i.e. $u_c=0$. This is also consistent with the fact that 
2-d random walks are recurrent, i.e. the interaction should be a relevant perturbation for any
$u>0$. It is difficult to obtain direct numerical evidence for this scenario, due to the very 
slow cross-over from the random walk behavior to the collapsed behavior, and due to the presence of strong corrections. In order to make any progress, we have to understand better these corrections.

In order to analyze the behaviour for very small $u$ more closely, let us define the quantity
\beq
   \Psi_t(u) = - {1\over u} \ln [R_t^2/t]              \label{psi}
\eeq
It is obviously well defined for $u\neq 0$, but it can be defined also for $u=0$ using l'H\^opital's
rule,
\beq
   \Psi_t(0) = - \lim_{u\to 0} {1\over u} \ln [R_t^2/t] = - {1\over t} {\partial R_t^2\over \partial u}.
\eeq
We used here the fact that $R_t^2 = t$ exactly for $u=0$. Numerically, $\Psi_t(0)$ can be estimated 
by a slight generalization of the reduced variance method discussed in subsection~\ref{variance}.
We simulate just ordinary random walks, but keeping track of the visited sites and calculating 
$\partial R _t^2 /\partial u$ using Eqs.~(2), (5), and (6). 

It is easily seen that $\Psi_t(u)$ is positive for all $u$. Plots of $\Psi_t(u)$ versus $t$, both for
positive {\it and for negative} values of $u$, are shown in Fig.~\ref{figPsi}.
Assume there is a collapse transition at $u=u_c$. We then expect that $\Psi_t(u)$ diverges as 
$\ln t$ for $u>u_c$ and $t\to\infty$, while it should stay bounded for $u<u_c$. More precisely, 
we expect that $\Psi_t(u) \sim const -a/t^\delta$ for $u<u_c$, where $\delta$ is another correction 
to scaling exponent. Plotting $\Psi_t(u)$ versus $t^{-\delta}$ should thus give straight lines 
converging to finite values for $t^{-\delta}\to 0$ if $u<u_c$, but upward bent curves diverging 
for $t^{-\delta}\to 0$ if $u>u_c$. One such plot, showing $\Psi_t(u)$ versus $t^{-0.22}$, is given in 
Fig.~\ref{figPsi2}. From this and similar plots with different exponents, we conclude that (i) the 
data are consistent with this scenario; (ii)
the critical point is at $u_c\approx 0$, most likely at $u_c=0$ exactly; (iii) the correction to 
scaling exponent is $\delta = 0.22\pm 0.05$; and (iv) at the critical point, $\Psi_t$ scales 
either as $\Psi_t(0) \sim \ln \ln t$ or $\Psi_t(0) \sim [\ln t]^\alpha$ with $0<\alpha \ll 1$. The 
former ($\Psi_t(0) \sim \ln \ln t$) seems preferred, but a clear distinction between these 
alternatives is not possible.

\begin{figure}
  \begin{center}
  \epsfig{file=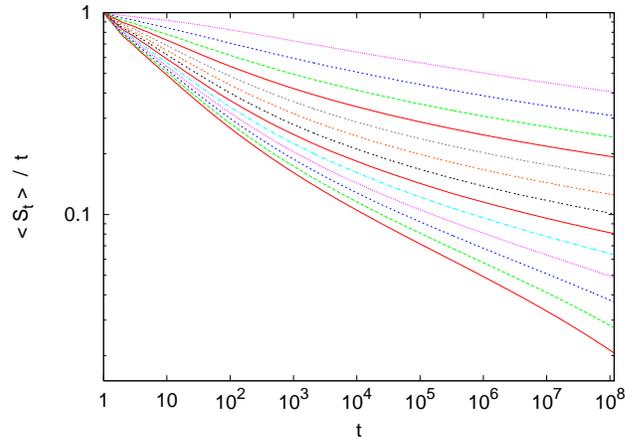, width=6.cm, angle=270}
  \caption{(color online) Same as Fig.~\ref{figS3}, but for $d = 2$. The curves correspond to $w = 0.2, 0.4, 0.6, \ldots 
     2.6$ (from top to bottom).}
  \label{figS2}
  \end{center}
\end{figure}

Studying $S_t$, the number of visited sites, is not very revealing. As seen from Fig.~\ref{figS2},
there is no value of $u$ for which the curve is straight.  $u\approx 0.7,\; w = e^u \approx 2$ 
yields the straightest curve in the large $t$ range $10^5 < t < 10^8$, but this is clearly not 
asymptotic, as the curves for larger $u$ indicate (they have not crossed over to asymptotic behavior 
and hence curve up for large $t$, although they finally curve down, for {\it very} large $t$).

\begin{figure}
  \begin{center}
  \epsfig{file=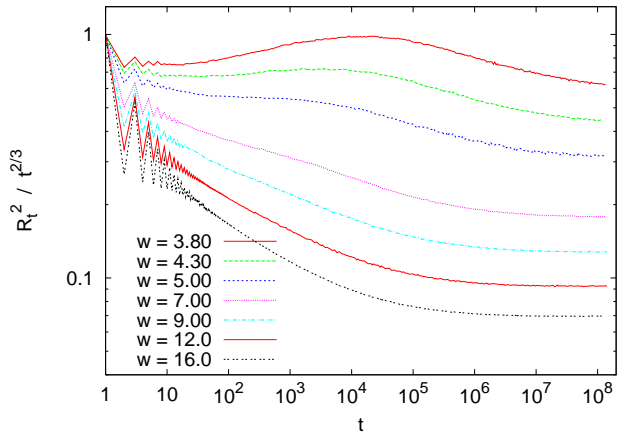, width=6.cm, angle=270}
  \caption{(color online) $R_t^2 / t^{2/3}$ for site TSATWs in 2 dimensions, for seven values of
      $u$ which are all larger than $u_c$. The curves seem to become horizontal as $t \to \infty$, with the asymptotic behavior appearing for smaller values of $t$ as $u$ increases.}
  \label{figR2s}
  \end{center}
\end{figure}

For coupling constants $u\gg 1$ one finds again that the prediction of the quasistatic 
approximation, $R_t \sim t^{1/3}$, is in excellent agreement with the data (see Fig.~\ref{figR2s}).
As in the 3-d case, corrections to this prediction are very large for small values of $u$, 
but they decrease quickly for $u\to\infty$. One would like of course to verify the 
quasistatic approximation for smaller $u$, but this seems at present impossible without going to lattice sizes beyond the reach of our normal computational resources.

\subsection{Bond TSATWs in $d = 3$}

\begin{figure}
  \begin{center}
  \epsfig{file=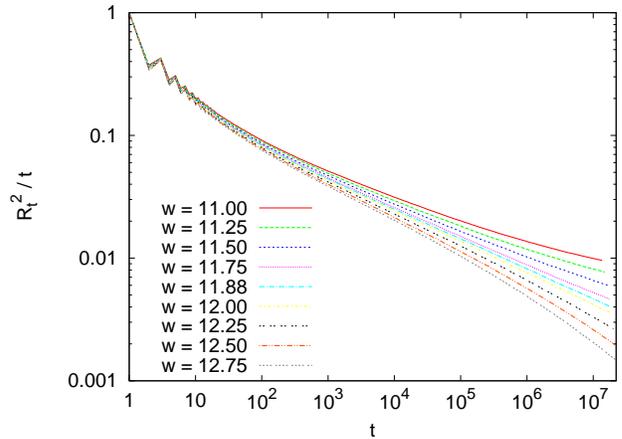, width=6.cm, angle=270}
  \caption{(color online) Average squared end-to-end distances for bond TSAWs in 3 dimensions, divided by 
      the number $t$ of steps. Each curve corresponds to a fixed value of $u$, with $u$ increasing 
      from top to bottom. The critical value $u=u_c = 2.475$ $(w_c \approx 11.88)$ corresponds to a straight curve in the limit 
      $t \to \infty$ whose slope is $2\nu_c - 1$. Here and in all subsequent figures, the curves are 
      labelled by $w = \exp(u)$. Statistical errors are comparable to the thickness of the curves.}
  \label{EFig2}
  \end{center}
\end{figure}

We now turn to the bond reinforced random walk in $d = 3$.  Results for the end-to-end distance $R_t^2$ are shown in Fig.~\ref{EFig2}.  The plot is superficially quite similar to Fig.~\ref{figR3}, with substantial corrections to scaling and the critical reinforcement $u_c$ occurring at much higher $u$.  This latter fact is unsurprising as bond-reinforcement is much more ``dilute" than its  site-reinforced cousin; consider that the equivalent of a visited site in the bond-reinforced model must have all six bonds visited in $d = 3$.  Analysis suggests $u_c = 2.475 \pm 0.003$ and $\nu_c = 0.380\pm 0.004$.  Note that $\nu_c$ is within error of the estimate $\nu_c = 0.378 \pm 0.004$ for the Site TSATW in $d = 3$.  This is the first piece of evidence that the bond and site models are in the same universality class.  

\begin{figure}
  \begin{center}
  \epsfig{file=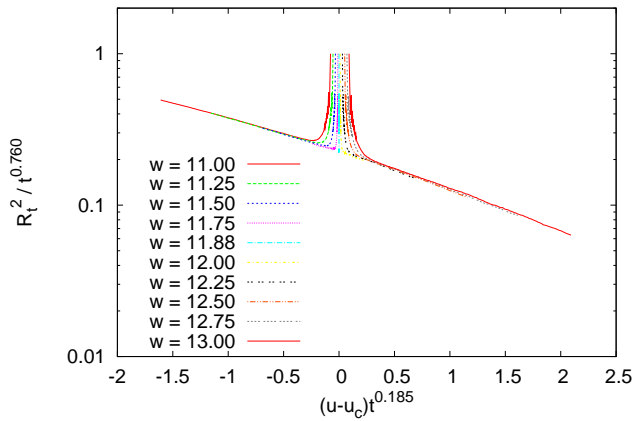, width=6.cm, angle=270}
  \caption{(color online) The same data as in Fig.~\ref{EFig2}, plotted now as $R_t^2/t^{2\nu_c}$ 
      versus $(u-u_c)t^\phi$, with $\nu_c = 0.380, u_c = 2.475$, and $\phi=0.185$.  Note that the critical exponents are within error bars of those used in the data collapse of Fig.~\ref{figcoll3}.}
  \label{EFig3}
  \end{center}
\end{figure}

In Fig.~\ref{EFig3} we show a data collapse with the same scaling ansatz Eq.(\ref{scale3});  $u_c = 2.475, \nu_c = 0.380$ and $\phi=0.185\pm 0.020$.  The critical exponents $\nu_c$ and $\phi$ are within error and identical, resp., to those for the Site TSATW in $d = 3$, see Fig.~\ref{figcoll3}, and the data collapse is if anything even better than that of Fig.~\ref{figcoll3}.  We see similar corrections to scaling at small $t$ and excellent collapse at large $t$, facilitated by our ability to simulate long walks (~$10^7$) due to the high value of $u_c$.  

Results for the average number of visited sites, $\langle S_t\rangle$ are not shown, but are similar to Fig.~\ref{figS3}, with substantial corrections to scaling.  We thus use the estimate of $u_c$ obtained from $R_t$, i.e. $u_c = 2.475 \pm 0.002$. The best estimate of the exponent $k_c$ (again made difficult by corrections to scaling) is $k_c = 0.970\pm 0.010$.  This is incompatible with the estimate $0.91\pm 0.01$ of \cite{ordemann2} but within the error of our estimate for the site-reinforced model, $k_c = 0.977 \pm 0.010$.  The leading correction to scaling exponent, defined as $\langle S_t\rangle = t^{k_c} [a +b/t^\Delta + o(t^{-\Delta})]$, is found to be $\Delta =0.25\pm 0.05$. This is within error of the estimate $\Delta =0.22\pm 0.03$ in the site case.

\begin{figure}
  \begin{center}
  \epsfig{file=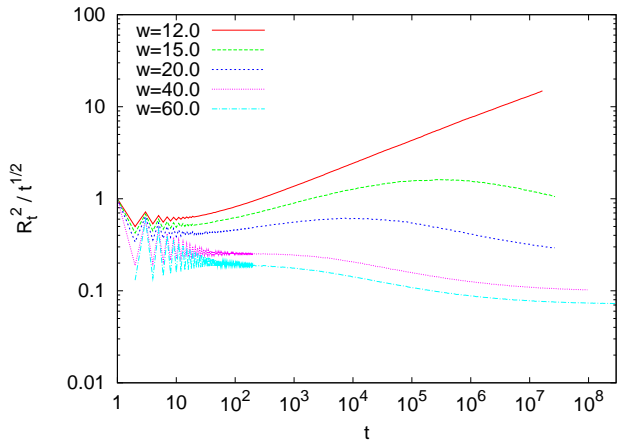, width=6.cm, angle=270}
  \caption{(color online) $R_t^2 / \sqrt{t}$ for bond TSAWs in 3 dimensions, for five values of 
      $u$ which are all larger than $u_c$.  The curves seem to become horizontal as $t \to \infty$, with the asymptotic behavior appearing for smaller values of $t$ as $u$ increases. This verifies that the quasistatic holds for bond as well as site reinforcement.}
  \label{EFig5}
  \end{center}
\end{figure}

Hence in all cases the estimates of the critical exponents for the site-reinforced model given by \cite{ordemann2} are excluded by our results as candidate exponents for the bond-reinforced model.  The estimates of all exponents for the bond-reinforced model agree (within error) with those we obtained for the site-reinforced case, see Sec.~\ref{site3}.  It is thus unsurprising that Fig.~\ref{EFig5} similarly verifies the quasistatic approximation $R_t\sim t^\nu$ with $\nu =0.25$ for $u > u_c$ in the large $t$ limit.  Crossover to the asymptotic behavior occurs at smaller $t$ as $u \to\infty$.  It is harder to verify the large $t$ limit for small $u > u_c$; we cannot run sufficiently long walks while limiting spurious self-intersection and hence reliably estimating $R_t$.

\subsection{Bond TSATWs in $d = 2$}

Our results for $d = 3$ indicate that the bond- and site-reinforced models are in the same universality class.  This implies that $u_c = 0$ for bond TSATW in $d = 2$.  To test this, we study small $u$ walks, which will cross over to the collapsed behavior only for large values of $t$.  This regime is even more difficult to study in the bond reinforced case, due to the dilute nature of the bond reinforcement.  The simulations used are as large as possible ($2^{32}$ sites and walks of $ \approx 10^8$ steps) but in many cases these walks just reach the beginning of what may be the scaling regime.

\begin{figure}
  \begin{center}
  \epsfig{file=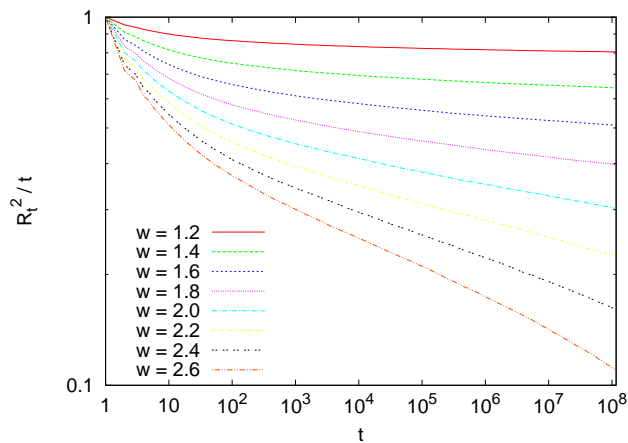, width=6.cm, angle=270}
  \caption{(color online) Same as Fig.~\ref{EFig2}, but for $d = 2$.}
  \label{EFig6}
  \end{center}
\end{figure}

\begin{figure}
  \begin{center}
  \epsfig{file=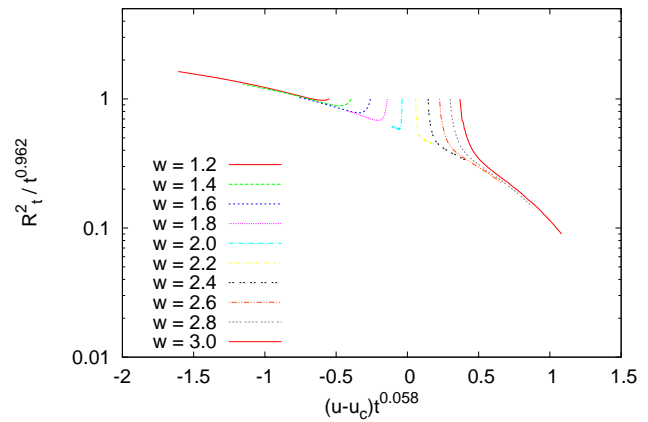, width=6.cm, angle=270}
  \epsfig{file=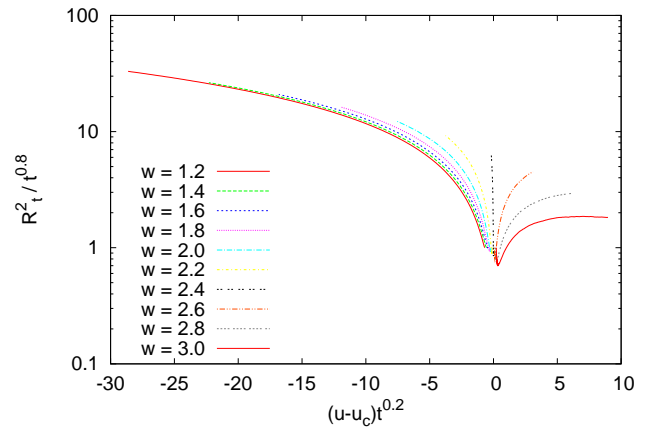, width=6.cm, angle=270}
  \caption{(color online) Attempted data collapse analogous to Fig.~\ref{EFig3}, but for the 
      data of Fig.~\ref{EFig6}. Parameters in panel (a) are $\nu_c = 0.481, u_c = 0.730$ $(w_c = 2.075)$, and 
      $\phi=0.058$, while panel (b) uses the values $\nu_c = 0.40, u_c = 0.88$ $(w_c = 2.411)$, $\phi=0.2$ 
      proposed in \cite{ordemann2}.  The latter collapse is completely unsatisfactory.} 
  \label{EFig7}
  \end{center}
\end{figure}

In Fig~\ref{EFig6} we show results for the end-to-end distance $R_t^2/t$.  As in the site-reinforced case for $d = 2$,  the curves fan out in Fig.~\ref{EFig6} for small $t$, with no apparent cross-over of the sort seen in Fig.~\ref{figR2} or Fig.~\ref{EFig2}.  An estimate of the critical reinforcement $u_c$ from these data is extremely difficult.  An attempted data collapse for the data of Fig.~\ref{EFig6}, using the scaling ansatz Eq.(\ref{scale3}) and 
optimized values $u_c = 0.73, \nu_c = 0.481$, and $\phi = 0.058$, is shown in Fig.~\ref{EFig7}(a).  While the data collapse acceptably for these values, the exponents proposed in \cite{ordemann2}, $u_c = 0.88, \nu_c = 0.40\pm 0.01$ and $\phi \approx 0.2$, can be ruled out, as a data collapse using these exponents is completely unsatisfactory, Fig.~\ref{EFig7}(b) . 

\begin{figure}
  \begin{center}
  \epsfig{file=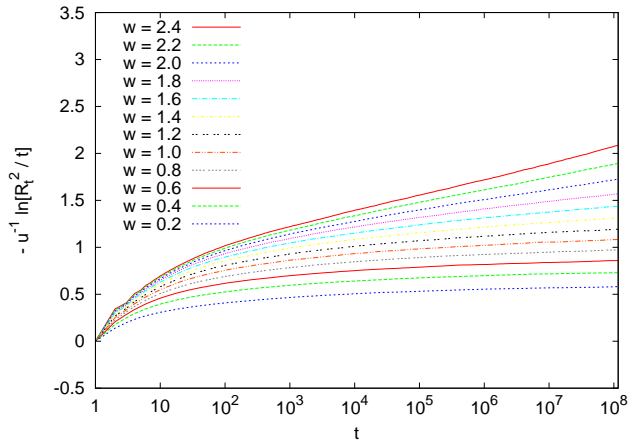, width=6.cm, angle=270}
  \caption{(color online) The function $\Psi_t(u)$ [see Eq.~(\ref{psi})] plotted against $t$ for
      various positive and negative values of $u$, including $u=0$. This figure again resembles a cross-over 
      plot as in Fig.~\ref{figPsi}, suggesting that $u=0$ $(w = 1)$ is a critical point for bond-reinforcement as well.}
  \label{EFig8}
  \end{center}
\end{figure}

As the estimated $\phi = 0.058$ is even smaller than that obtained for the site-reinforced model in $d = 2$ (where $\phi = 0.085$) and as $\nu_c = 0.481$ is very close to the random walk value $\nu = 1/2$, we argue that there is no phase transition for $u_c > 0$ in the bond-reinforced model, either. In particular, similar heuristic arguments about the recurrence of 2-d random walks suggest again that any non-zero reinforcement is a relevant perturbation.  Hence we study again the function $\Psi_t(u)$ [Eq.~(\ref{psi})] for $u > 0$, $u = 0$, and $u < 0$ (see Fig.~\ref{EFig8}). As for the site case (subsection \ref{site2}), plotting $\Psi_t(u)$ against $1/t^\delta$ with different exponents $\delta$ reveals the detailed asymptotic behavior. Such plots (not shown here) indicate that $u_c = 0$, and that the correction to scaling exponent in the uncollapsed phase $u<u_c$ is $\delta = 0.20 \pm 0.05$, well within error of the estimated $\delta = 0.22 \pm 0.05$ of the site case.

As was the case for the site-reinforced model in $d = 2$, it is not particularly illuminating to study the number of visited sites $S_t$.  The corrections to scaling are even larger, and hence it is impossible to estimate the correct scaling exponent from these data.

\begin{figure}
  \begin{center}
  \epsfig{file=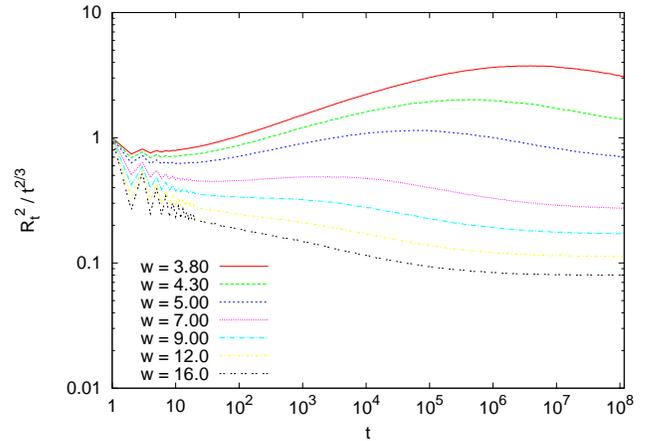, width=6.cm, angle=270}
  \caption{(color online) $R_t^2 / t^{2/3}$ for bond TSAWs in 2 dimensions, for seven values of
      $u$ which are all larger than $u_c$. The curves seem to become horizontal for $t \to \infty$
      and $u \to \infty$.}
  \label{EFig10}
  \end{center}
\end{figure}

For coupling constants $u\gg 0$ one finds that the prediction of the quasistatic 
approximation, $R_t \sim t^{1/3}$, is in excellent agreement with the data (see Fig.~\ref{EFig10}).  Corrections to this prediction are very large for small values of $u$, 
but become irrelevant as $u \to \infty$, as is already apparent at $w = e^{u}=16$.  Familiar limitations to accessible lattice size make the quasistatic approximation impossible to verify for smaller $u$, where crossover to the asymptotic behavior takes place at very large $t$.  Settling the validity of the quasistatic approximation in the small $u$ regime will in all likelihood require the development and application of appropriate analytical methods.

\section{Discussion and Outlook}
Despite its simplicity, the once-reinforced site variant of the True Self Attracting Walk (TSATW) has generated considerable controversy since its original statement by Sapozhnikov \cite{sapo}.  In this paper we have used a combination of careful high-statistics simulations and heuristic arguments to attempt a resolution of many of these disputes.  

In $d = 3$ we confirm the existence of a phase transition from random walk-like to collapsed behavior for finite reinforcement $u_c$.  Our simulations provide overwhelming evidence for rejecting the proposed $u_c$ and scaling exponents of \cite{ordemann1, ordemann2}.  We find  $u_c = 1.831\pm 0.002$ and $\nu_c = 0.378\pm 0.004$, with the crossover exponent $\phi = 0.185 \pm 0.020$.  In addition we verify the quasistatic approximation $R_t^2 \sim \sqrt{t}$ for large $t$ and $u$.  

In $d = 2$ we argue that there is no phase transition at any finite reinforcement $u_c$.  For any $u>0$ the walks go to a collapsed phase, and the ``critical behavior" at $u_c = 0$ is simply that of a random walk.  The quasistatic approximation $R_t^2 \sim t^{2/3}$ is also verified for large $t$ and $u$. 

In addition to the site-reinforced TSATW, we studied the bond-reinforced variant and found evidence that despite the underlying mathematical differences (and related difficulties) bond-reinforced TSATW is in the same universality class as site-reinforced TSATW.  In $d = 3$ we found a phase transition at finite reinforcement $u_c = 2.475 \pm .003$ and scaling behavior extremely similar to the site-reinforced model, with similar success of the quasistatic approximation.  In $d = 2$ we found evidence of a phase transition at $u_c = 0$ although the evidence here is somewhat weaker due to the long time needed to cross over to the collapsed behavior and the memory limitations imposed by the extremely large lattices needed to minimize spurious self-intersection.  

An obvious limitation of our work is the lack of an analytical understanding as to why the transition to a collapsed phase should occur at any finite reinforcement in the site- and bond- models in $d = 2$.  In the mathematics literature, the once-reinforced ERRW has been studied by mapping it to a diffusion with a drift term (directed inward) at the boundary \cite{pemantle, dav96}.  As far as we know this technique has only been applied in $d = 1$ and is even in this case of considerable technical difficulty.  There are also techniques mapping the stochastic process to a (deterministic) dynamical system, the so-called ``stochastic approximation", which is largely unknown to the physics literature \cite{pemantle, pemantle1}.  This suggests that some sensible map to a continuous process or a dynamical system might enable an analytic proof of $u_c = 0$ in one or both of the variants of once-reinforced TSATW in $d = 2$. 

More generally, the universality result we propose for site and bond TSATW in $d = 3$ and $d = 2$ suggests the possibility of a deep dialogue between the statistical physics and probability literatures.  The perspective of statistical physics generates different questions (with respect to phase transitions, critical behavior, and universality) that complement the rigorous results derived within the probability community.  Furthermore, the probability literature as reviewed in \cite{pemantle} contains an enormous number of unexplored models for random walks with reinforcement. It is also clear that these walk processes are specific instances of a general study of random processes with reinforcement, with many applications in the biological and social as well as physical sciences \cite{pemantle}.  The statistical physics of such models remains an almost entirely open question.  

We end on a cautionary note about the use of simulation in these problems.  As pointed out by Pemantle \cite{pemantle1}, the convergence times for some random processes with reinforcement can be astronomical; the Friedman urn, for example, does not reach its asymptotic behavior until a googol updates or more.  This suggests that in some cases the high statistics simulations that would be applied by statistical physicists may only be probing the transient behavior of such models.  While the transient behavior has its own intrinsic interest, we suggest that a dialogue between the two fields would do much to drive research in mutually beneficial directions--while avoiding pitfalls along the way.   

Acknowledgment: J.G.F. and P.G. are supported by iCORE.


\begin{thebibliography}{99}
\bibitem{deGennes} P.-G. de Gennes, {\it Scaling Cpts in Polymer Physics} (Cornell University Press,
   Ithaca, 1979).
\bibitem{Amit} D. Amit, G. Parisi, and L. Peliti, Phys. Rev. B {\bf 27}, 1635 (1983).
\bibitem{domb} C. Domb and G.S. Joyce, J. Phys. C {\bf 5}, 956 (1972).
\bibitem{donsker} M.D. Donsker and S.R.S Varadhan, Comm. Pure Appl. Math. {\bf 28}, 525 (1975); 
   {\bf 32}, 721 (1979).
\bibitem{mehra} V. Mehra and P. Grassberger, Physica D {\bf 168}, 244 (2002).
\bibitem{sapo} V.B. Sapozhnikov, J. Phys. A {\bf 27}, L151 (1994); {\bf 27}, 3935 (1998).
\bibitem{reis} F.D.A. Aaroro Reis, J. Phys. A {\bf 28}, 3851 (1995).
\bibitem{prasad} M.A. Prasad, D.P. Bhatta, and D. Arora, J. Phys. A {\bf 29}, 3037 (1996).
\bibitem{lee} J.W. Lee, J. Phys. A {\bf 31}, 3929 (1998).
\bibitem{ordemann1} A. Ordemann, G. Berkolaiko, S. Havlin, and A. Bunde, Phys. Rev. E {\bf 61}, R1005 (2000).
\bibitem{ordemann2} A. Ordemann, E. Tomer, G. Berkolaiko, S. Havlin, and A. Bunde,
    Phys. Rev. E {\bf 64}, 046117 (2001).
\bibitem{dalmaroni} A. Jim{\'e}nez-Dalmaroni and H. Hinrichsen, Phys. Rev. E {\bf 68}, 036103 (2003).
\bibitem{pemantle} R. Pemantle, Probability Surveys {\bf 4}, 1 (2007); arXiv:math.PR/0610076v2 (2007).
\bibitem{pemantle1} R. Pemantle, Random processes with reinforcement, http://citeseer.ist.psu.edu/617028.html.  Preprint (2001).  
\bibitem{trails} A. Guttman, J. Phys. A {\bf 18}, 567 (1985).
\bibitem{dia88} P. Diaconis,  In J. Bernardo, M. de Groot, D. Lindley, and A. Smith, eds., {\it Bayesian Statistics}, pp. 111-125.  (Oxford Univresity Press, Oxford, 1988).
\bibitem{kr99} M. Keane and S. Rolles, In {\it Infinite Dimensional Stochastic Analysis} volume 52 of {\it Verhandelingen, Afdeling Natuurkunde.  Eerste Reeks.  Koninklijke Nederlandse Akademie van Wetenschappen}, pp. 217-234.  (R. Neth. Acad. Arts Sci., Amsterdam, 1999).  
\bibitem{tar04} P. Tarr{\`e}s, Ann. Probab. {\bf 32}, 2650 (2004).
\bibitem{pv99} R. Pemantle and S. Volkov,  Ann. Probab. {\bf 27}, 1368 (1999).
\bibitem{vol1} S. Volkov, Ann. Probab. {\bf 29}, 66 (2001). 
\bibitem{dkl02} R. Durrett, H. Kesten, and V. Limic, Prob. Thoer. Rel. Fields {\bf 122}, 567 (2002).
\bibitem{die05} J. Die,  Statistics Probab. Lett. {\bf 73}, 115 (2005). 
\bibitem{sel06} T. Sellke, Elec. J. Prob. {\bf 11}, 301 (2006).  
\bibitem{sellke} T. Sellke, Reinforced random walks on the $d$-dimensional lattice.  Preprint (1994).  
\bibitem{torney} D.C. Torney, J. Stat. Phys. {\bf 44}, 49 (1986).
\bibitem{derkachov} S.E. Derkachov, J. Honkonen and A.N. Vasil'iev, J. Phys. A: Math. Gen. {\bf 23}, 2479 (1990).
\bibitem{dav96} B. Davis, Ann. Probab. {\bf 24}, 2007 (1996).  
\end{thebibliography}
\end{document}